# Stepping beyond your comfort zone: Diffusion-based network analytics for knowledge trajectory recommendation

Yi Zhang[*], Mengjia Wu, Jie Lu

Australian Artificial Intelligence Institute, Faculty of Engineering and Information Technology, University of Technology Sydney, Australia

**Email:** Yi.Zhang@uts.edu.au (corresponding author); Mengjia.Wu@uts.edu.au; Jie.Lu@uts.edu.au

**Abstract**

Interest in tracing the research interests of scientific researchers is rising, and particularly that of predicting a researcher's knowledge trajectories beyond their current foci into potential inter-/cross-/multi-disciplinary interactions. Hence, in this study, we present a method of diffusion-based network analytics for knowledge trajectory recommendation. The method begins by constructing a heterogeneous bibliometric network consisting of a co-topic layer and a co-authorship layer. A novel link prediction approach with a diffusion strategy is then used to reflect real-world academic activity, such as knowledge sharing between co-authors or diffusing between similar research topics. This strategy differentiates the interactions occurring between homogeneous and heterogeneous nodes and weights the strengths of these interactions. Two sets of experiments – one with a local dataset and another with a global dataset – demonstrate that the proposed method is prior to selected baselines. In addition, to further examine the reliability of our method, we conducted a case study on recommending knowledge trajectories of selected information scientists and their research groups. The results demonstrate the empirical insights our method yields for individual researchers, communities, and research institutions in the information science discipline.

**Introduction**

Dating back to the early 1980s, the continuous and discontinuous changes in a technology along with its developmental progress drew attention from Dosi (1982). He defined the continuous changes as technological trajectories, highlighting the cumulative process of technical advances is in line with an established routine. Discovering technological opportunities along its trajectories then became a breakthrough for promoting technological innovation (Porter & Detampel, 1995; Souitaris, 2002). In assembling scientific research and technological development within scenarios of knowledge, *knowledge trajectories* refer to how knowledge is integrated and differentiated within dynamic processes of change (Barley et al., 2018). Ways of understanding these trajectories in a dynamic setting have been receiving attention for decades.

The innovation literature highlights the correlations between knowledge trajectories with internal and external innovation factors (Castellacci, 2008). It also examines its relationships with innovation outcomes (Furman & Teodoridis, 2020; Majchrzak & Malhotra, 2016), and the cross-impacts between multiple trajectories (Ciarli et al., 2021). In the field of information science, science maps and data visualisation tools have created great opportunities to map knowledge trajectories. As a result, diverse bibliometric indicators have been revealed, such as citation linkages and semantic similarities. Additionally, innovation paradigms like technology roadmapping have come to the fore (Fu et al., 2019; Lee et al., 2009; Mina et al., 2007; Zhang et al., 2013). Despite a keen interest in these endeavours, predicting the future trajectories of knowledge still requires heavy human intervention, and usually these interventions involve strategies of combining quantitative and qualitative approaches (Robinson et al., 2013; Zhang et al., 2016).

Techniques in complex network analytics (Borgatti et al., 2009; Palla et al., 2005) have also raised new ideas and increased our technical capability for further analysing the topological structures of science maps. As a consequence, many hidden relationships behind simple visualisations have been explored (Zhang, Wu, Miao, et al., 2021). Examinations of the network features in science maps (e.g., small world and scale free) endorse the assumption that a science map is the counterpart of a bibliometric network (Hung & Wang, 2010; Liu et al., 2012). This opens a door for information scientists to facilitate network analytics in broad information studies that, say, measure research impacts (Yan & Ding, 2009) or technological innovation (Leydesdorff & Rafols, 2011; Yeo et al., 2015).

Significantly, link prediction, with its core assumption that two unconnected nodes will be linked in future if common neighbours exist (Liben-Nowell & Kleinberg, 2007), surprisingly coincides with a well-recognised observation in the innovation literature – that is, that innovation is the recombination of established knowledge (Fleming, 2001). Given a homogeneous bibliometric network, such as a co-term, co-citation, or co-authorship network, some pilot studies have used link prediction to recommend potential collaborators (Yan & Guns, 2014) or to predict emerging technologies (Érdi et al., 2013; Zhou et al., 2019). We argue that a heterogeneous network with multiple bibliometric indicators may form a better foundation for comprehensively understanding the interactions between both homogeneous indicators (e.g., collaboration) and heterogeneous indicators (e.g., shared research interests between researchers). Thus, developing an approach of link prediction for analysing heterogenous bibliometric networks could be beneficial for predicting the potential interactions between multiple bibliometric entities. However, undertaking such a study is a challenge since algorithmically measuring the interactions between heterogeneous and homogeneous nodes to truly reflect academic activities in the real world is a difficult task.

Following the definition given by Barley et al. (2018), in our research, we consider knowledge trajectories as the changing process of scientific and technological topics. When zooming in to an individual's knowledge, two types of knowledge trajectories exist: (1) the historical trajectory of a researcher's knowledge, which can be profiled by mapping his/her research topics in a timeline; and (2) future knowledge trajectories that can either be vertical or horizontal extensions of existing knowledge (or both). Vertical extensions requires knowledge acceleration in relation to existing topics, while a horizontal extension may require knowledge recombination from inter-/cross-/multi-disciplines (Jones, 2021). This study focuses on recommending knowledge trajectories for a target researcher in terms of horizontal extensions of his/her knowledge base. Notably, this extension may require a long-term research vision and a shift beyond his/her current comfort zone. As such, this paper proposes a novel method for analysing heterogeneous bibliometric networks and recommending knowledge trajectories to target researchers.

The method begins by constructing a heterogeneous bibliometric network with a co-topic layer and a co-authorship layer. The interactions between homogeneous and heterogeneous nodes in this network are then predicted by a model of diffusion-based link prediction that relies on network-based inference (NBI) (Zhou et al., 2007). However, the scope of the inference is extended from a bipartite network to a bi-layer network. This is because bipartite network analytics ignore knowledge diffusion between the same set of nodes, yet as argued above, such interactions may reflect significant academic activity. Seven baselines are assembled for validation measurements: four link prediction techniques, two recommender systems, and one specific approach that uses embedding techniques to construct a semantic layer rather than a co-topic layer. We assembled two datasets for testing: a local dataset that contains 11,399 journal articles from the information science literature, and a global dataset comprising the full set of the Digital Bibliography & Library Project (DBLP) database, covering 4.89 million research articles in computer science. The validation demonstrates the reliability of our method in recommending knowledge trajectories.

Beyond our experiments, we also conducted a case study using the local dataset. This study explores empirical insights for information scientists by profiling research groups, core topics, and possible trajectories for specific research groups to horizontally expand research foci. These insights are useful for providing empirical decision support to individual researchers, research institutions, and funding agencies in the information science discipline.

**Related work**

*Bibliometric network analytics*

Science maps provide solutions of visualising bibliometric entities and their relationships via nodes and edges in a network(Garfield, 2004; Leydesdorff, 1987). Previous studies have either developed mapping capabilities for enhancing its representation (Rafols et al., 2010; Rotolo et al., 2017; Suominen & Toivanen, 2016) or applied science maps to broad practical scenarios (Daim et al., 2017; Petralia, 2020). Further, as the impact of complex network analytics widened from applied physics to computer science and onwards to the social sciences (Borgatti et al., 2009), information scientists leapt at the opportunity to incorporate social network analytics into science maps (Björneborn, 2004). In this paper, we used the term *bibliometric networks* to describe science maps. Our particular focus is on their topological structures.

Previous studies incorporating bibliometric network analytics have: 1) used topological indicators, such as centrality, to identify key nodes and determine their actual meanings – e.g., influential researchers in a co-authorship network (Li et al., 2013; Yan & Ding, 2009); 2) used topology-based approaches, such as community detection and link prediction, to recognise specific behaviours, relationships, and patterns, e.g., collaborations (Yan & Guns, 2014), disciplinary interactions (Huang et al., 2020), or problem-solving patterns (Zhang, Wu, Hu, et al., 2021); 3) connected bibliometric networks with a broad scenario of innovation paradigms, e.g., technology roadmaps (Jeong et al., 2021) and technology opportunity analysis (Park & Yoon, 2018; Ren & Zhao, 2021).

Very few studies use heterogenous bibliometric networks, but there are two worth highlighting. Aiming to understand the collaborative/citing patterns of academic researchers, Ding (2011) applied an approach to a citation network and a co-authorship network that incorporated topic models with a random walk approach. Compared to typical network analytics, this work creatively embedded topic models with two bibliometric indicators. The other study is our adventure in applying heterogenous bibliometric networks for measuring emerging general-purpose technologies (Zhang, Wu, Miao, et al., 2021). In this pilot study, we constructed a heterogenous network with a co-authorship layer and a co-term layer for link prediction. However, it did not differentiate between the interactions of homogeneous nodes and heterogeneous nodes but just treated all edges equally in resource allocation.

*Scholarly recommendation*

The task of recommending knowledge trajectories is in line with scholarly recommendation as well. Scholarly recommendation inherits the basic settings of recommender systems, i.e., recommending suitable items to target users by analysing their either explicit or implicit relationships (Lu et al., 2015), but instead of typical user-item pairs, it mainly targets academic researchers, recommending academic outlets (Alhoori & Furuta, 2017; Pera & Ng, 2011) and counterparts (e.g., collaborators, reviewers, and supervisors) (Chughtai et al., 2020; Liu et al., 2018; Rahdari et al., 2020).

Despite novel recommender systems for complicated settings, such as cross-domain recommendation (Tang et al., 2012) and multi-relational recommendation (Mao et al., 2016; Xu et al., 2019), popular real-world applications of recommender systems still heavily rely on two traditional approaches: content-based and collaborative filtering-based approaches. Since recommender systems require typical data features for validation, such as tags and rates, scholarly recommendation is mainly used to support well-labelled digital libraries (Sinha et al., 2015) and reference management tools (Alhoori & Furuta, 2017). However, bibliometric databases usually contain extremely limited or rare labels for tagging/rating scholarly items, and, given the task of recommending knowledge trajectories, the absence of golden standards adds further challenges.

**Methodology: Diffusion-based network analytics**

This paper is on the trail of intelligent bibliometrics (Zhang et al., 2020) – developing computational models elaborating artificial intelligence and data science techniques with bibliometric indicators for handling issues in ST&I studies. Specifically, this paper is to develop a method of diffusion-based network analytics to recommend knowledge trajectories for target researchers. The research framework is given in Figure 1. It includes three phases: data pre-processing, bi-layer network construction, and diffusion-based prediction.

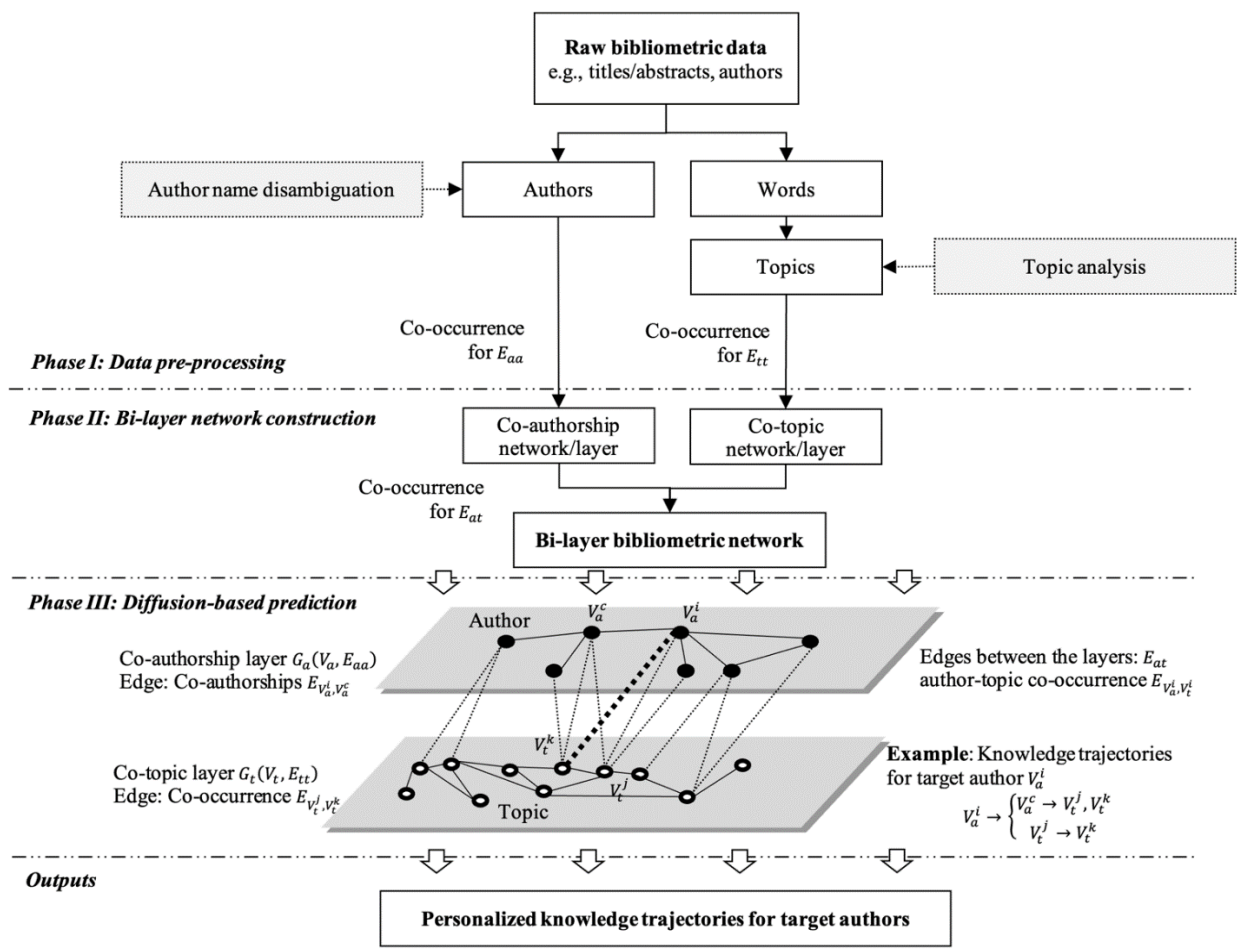

Figure 1. Research framework of diffusion-based network analytics for recommending knowledge trajectories

*Phase I: Data pre-processing*

The proposed method targets bibliometric data, such as scientific documents, patents, and academic proposals, and more specifically two fields of bibliographical information: the combined text of an article's title and abstract and its authorship. Thus, two pre-processing tasks are required:

- *Pre-processing author names*: Author names in raw bibliometric data may vary hugely, appearing as, say, "Eugene Garfield", "Garfield, Eugene", "Garfield, E", and "E Garfield". Author name disambiguation is therefore required to consolidate variations and remove unidentified names.
- *Topic extraction and representation*: Targeting the titles and abstracts, topic extraction or topic modelling is conducted to identify and label topics from the corpus.

*Phase II: Bi-layer network construction*

Our bi-layer network consists of a co-authorship layer and a co-topic layer. Briefly, on the co-authorship/co-topic layer, a node represents an author/topic, and an edge represents the co-occurrence between connected authors/topics, which is weighted by the co-occurrent frequency. Notably, one of the most recent studies on topic taxonomy construction (Shang et al., 2020) raised a major drawback of embedding in local text data, that is, word embedding techniques cannot effectively distinguish highly coupled words in a specific domain. Given that, we decided to keep the co-occurrence on the co-topic layer in our default setting. However, in cases with broad disciplinary interactions, the semantic similarities between topics may serve as an option for weighting the edges in the topic layer.

Referring to Figure 1, the bi-layer network is described as: $G\{G_a(V_a, E_{aa}), G_t(V_t, E_{tt}), E_{at}\}$, where $G_a(V_a, E_{aa})$ represents a co-authorship layer with the author nodes $V_a$ and the edges $E_{aa}$. $G_t(V_t, E_{tt})$ represents a co-topic layer, with topic nodes $V_t$ and edges $E_{tt}$; and $E_{at}$ represents the edges between the two layers.

Let $|E|$ represent the weight of an edge $E$. Let $V_a^i$ and $V_a^c$ be nodes in the co-authorship layer and let $V_t^k$ and $V_t^j$ be nodes on the co-topic layer. Then the weights of the three types of edges can be represented as:

$$\left|E_{V_a^i, V_a^c}\right| = \theta_{V_a^i, V_a^c} \tag{1}$$

$$\left|E_{V_t^k, V_t^j}\right| = \varphi_{V_t^k, V_t^j} \tag{2}$$

$$\left|E_{V_a^i, V_t^j}\right| = \mu_{V_a^i, V_t^j} \tag{3}$$

where $\theta_{V_a^i, V_a^c}$, $\varphi_{V_t^k, V_t^j}$, and $\mu_{V_a^i, V_t^j}$ are constants, representing the co-occurrent frequency between connected nodes.

*Phase III: Diffusion-based prediction*

Resource allocation is a key concept in diffusion-based prediction, which is rooted in the NBI approach proposed by Zhou et al. (2007). NBI was designed for a bipartite network, which refers to a network consisting of two sets of nodes. In NBI, resources diffuse only between nodes from different sets, where their common neighbours serve as transmitters to distribute resources (Ou et al., 2007). Thus, given a target node from which the initial resources are created, the amount of resources that a node can eventually receive becomes an indicator that quantifies the closeness of the connections between that node and the target node. Initially, these connections are considered to be a user's potential preference for an item (Zhou et al., 2007). The NBI approach then further became the core of a resource allocation-based link prediction approach (Zhou et al., 2009). In our pilot studies, we applied this approach as a way to recommend a researcher's potential research interests if the bipartite consisted of authors and terms. And, if the bipartite consisted of terms and terms, we deemed it represented a research topic's future direction(s) (Zhang, Wang, et al., 2018). We also weighted this index according to the strength of these connections and applied it to predict potential connections (Zhang, Wu, Miao, et al., 2021). However, when applying this typical setting of a bipartite network to a bibliometric scenario, two issues may be encountered:

- In a bibliometric bipartite network, the interactions between homogeneous nodes (i.e., nodes from the same set) may contain crucial and unignorable information for measuring academic activity. For example, collaboration can drive knowledge diffusion (Singh, 2005), and numerous terminological variations and synonyms, i.e., interactions between topics, can be a key attribute of emerging technologies observed in bibliometric texts (Zhang et al., 2014).
- In terms of heterogeneous networks, differentiating the diffusion strategies between homogeneous nodes and between heterogenous nodes may reflect actual meanings, but there is rare literature touching on a diffusion strategy for comprehensively describing academic activities in a heterogeneous bibliometric network.

These two issues inspired the design of our proposed method. They are why we extended the scope of the resource allocation from a bipartite network to a bi-layer network and are why the network's topological structure is analysed from the perspective of heterogeneous network analytics. For this reason, our focus is on analysing three types of edges in the bi-layer network $G$: $E_{aa}$, $E_{tt}$, and $E_{at}$. Additionally, we redesigned the resource diffusion strategy to predict the potential edges $E_{at}$ between a target author and topics. The algorithm of the diffusion-based prediction for a target author $V_a^i$ is described as follows:

**Step 1** – *Diffusion via author-topic edges $V_a^i \rightarrow V_t^j$*: This is a typical process of resource allocation designed by the NBI approach but with a weighting solution added to the diffusion strategy. If an author $V_a^i$ holds the initial resources $r(V_a^i)$, a portion of those resources will spread to the connected topics $V_t^j$. The resource $f(V_t^j)$ that topic $V_t^j$ will receive from author $V_a^i$ can be calculated as:

$$f(V_t^j) = \frac{\left|E_{V_a^i,V_t^j}\right|}{\Sigma_{E_{V_a^i,V_t^p}\neq 0}\left|E_{V_a^i,V_t^p}\right|} r(V_a^i) \qquad (4)$$

**Step 2** – *Diffusion via author-author edges $V_a^i \rightarrow V_a^c$*: Assuming academic researchers are willing to share knowledge with their co-authors, author $V_a^i$ will 'copy' the same amount of initial resources $r(V_a^i)$ and spread them to connected authors $V_a^c$ (i.e., co-authors) based on their co-authorship strengths. The resources $f(V_a^c)$ that author $V_a^c$ will receive from author $V_a^i$ can be calculated as:

$$f(V_a^c) = \frac{\left|E_{V_a^i,V_a^c}\right|}{\Sigma_{E_{V_a^i,V_a^q}\neq 0}\left|E_{V_a^i,V_a^q}\right|} r(V_a^i) \qquad (5)$$

**Step 3** – *Diffusion via topic-topic edges $V_t^j \rightarrow V_t^k$*: Assuming the co-occurrence between research topics indicates pairwise knowledge sharing, a topic $V_t^j$ will spread the resources it has acquired to connected topics based on their co-occurrence strengths. The resource $f(V_t^k, V_t^j)$ that topic $V_t^k$ will receive from topic $V_t^j$ and the total resource $f_t(V_t^k)$ that topic $V_t^k$ will receive from connected topics can be calculated as:

$$f(V_t^k, V_t^j) = \frac{\left|E_{V_t^j,V_t^k}\right|}{\Sigma_{E_{V_t^j,V_t^p}\neq 0}\left|E_{V_t^j,V_t^p}\right|} f(V_t^j) \qquad (6)$$

$$f_t(V_t^k) = \Sigma_{E_{V_t^k,V_t^p}\neq 0} f(V_t^k, V_t^p) \qquad (7)$$

**Step 4** – *Diffusion via author-topic edges $V_a^c \rightarrow V_t^k$*: Repeat step 1 but for the target author's co-authors $V_a^c$, who will diffuse their resources to connected topics as well. Thus, the resources $f(V_t^k, V_a^c)$ that topic $V_t^k$ will receive from the author $V_a^c$ and the total resources $f_a(V_t^k)$ that topic $V_t^k$ will receive from the target author's co-authors can be calculated as:

$$f(V_t^k, V_a^c) = \frac{|E_{V_a^c, V_t^k}|}{\Sigma_{E_{V_a^c, V_t^p} \neq 0} |E_{V_a^c, V_t^p}|} f(V_a^c, V_a^i) \tag{8}$$

$$f_a(V_t^k) = \Sigma_{E_{V_a^q, V_t^k} \neq 0} f(V_t^k, V_a^q) \tag{9}$$

**Step 5** – *Resource finalisation*: Since the objective of this prediction is to recommend new research topics to a target author – that is, topics beyond their comfort zone – our focus is solely on topics unconnected to the target author, i.e., topics $V_t^k$. Thus, the final resource $f(V_t^k)$ that topic $V_t^k$ will receive can be calculated as:

$$f(V_t^k) = f_t(V_t^k) + f_a(V_t^k) \tag{10}$$

**Outputs** – *Ranking and personalised recommendation*: The output of the proposed method is a ranking list $R$ containing a list of the target author's $V_a^i$ unconnected topics $V_t^k$, ranked by their final resource $f(V_t^k)$. This list is, in fact, personalised, since this list $R$ is generated based on this target author's co-authorships and his/her own research topics. Obviously, such a list of recommendations will differ case by case.

Steps 1 and 2 describe a scenario where authors are open to share knowledge with their co-authors, and Step 3 reveals that co-occurred topics can act as a mediator for knowledge sharing. Both scenarios are designed to maximise the simulation of knowledge diffusion in real-world academic activities. Also, research topics beyond the target author's current foci are recommended, which may be that author's future knowledge trajectories.

*Validation measurements*

When considering bibliometric data containing time stamps, i.e., publication year, the data can be divided into two sub-datasets and used for validation. We divided our dataset into the most recent 5 years and used that data for testing, while the remaining 'old' data was used for training.

Considering the core of the proposed method follows the main logic of link prediction – i.e., Phase III and the diffusion-based prediction – we selected five widely used link prediction baselines in the bibliometric literature for comparison:

- *Jaccard Coefficient* (JC): A common neighbour (CN)-based algorithm that calculates the proportion of common neighbours between two unlinked nodes.
- *Adamic-Adar Index* (AA): A CN-based algorithm that assigns more weights to common neighbours with smaller degrees (Adamic & Adar, 2003).
- *Preferential Attachment* (PA): An algorithm assuming that the more connected a node is, the more likely it is to receive new links (Newman, 2001).
- *Resource Allocation* (RA): A CN-based algorithm that allocates resources according to the degree of their CNs (Zhou et al., 2009).
- *Weighted Resource Allocation* (Weighted RA): A refined RA algorithm that uses a weighted index to involve edge weights (Zhang, Wu, Miao, et al., 2021).

Further, since the proposed method is designed to recommend research topics to a target researcher, which is the work of a typical recommender system (Lu et al., 2015), we also introduced two recommender systems algorithms as baselines:

- *Content-based recommender systems* (Content): Recommendations are based on the similarity between a target author's topics of interest and others. The key assumption is that an author will be interested in similar areas to their current focus.
- *Collaborative filtering-based recommender systems* (CF): Recommendations are based on the research topics of a target author's co-authors. The key assumption is that an author will be interested in the research topics of their co-authors.

We then conducted a comparative experiment (Semantic Diffusion for short) to examine whether the drawback of embedding in local text data (Shang et al., 2020) existed in our local bibliometric dataset. Specifically, we: (1) represented each field of study (FoS) tag as a topic vector of assembled word vectors generated using the Word2Vec approach (Mikolov et al., 2013); (2) measured the cosine similarities between the topic vectors and constructed a semantic layer to replace the co-topic layer; and (3) applied diffusion-based prediction to this new bi-layer network to generate recommendations.

In terms of validation measures, we calculated the receiver operating characteristics (ROC) and area under curve (AUC) – both widely recognised indicators in machine learning and network science (Bradley, 1997; Fawcett, 2006). Briefly, given two dimensions in a ROC curve, a true-positive result gains a plot in the Y axis while a false-positive result gains a plot in the X axis. Then, the AUC value, which is the area under the ROC curve, is used to validate the measurement. Given an interval of [0, 1], the larger the AUC value, the better the approach performs.

**Results**

*Data description and pre-processing*

The DBLP database[1] is well known for covering research articles published in major computer science (CS) journals and proceedings, highlighting the CS community's particular recognition in not only high-quality journals but also reputable conferences. With the aid of the open data platform AMiner (Tang et al., 2008), we collected 4,894,081 articles indexed by DBLP on April 9, 2020 and before – i.e., the DBLP-Citation-network v12.

We chose AMiner since its released data have already been pre-processed and stored in knowledge graphs. Specifically: (1) Author names have already been disambiguated (Tang et al., 2011) with 4,398,138 distinctive authors identified in the collected dataset; and (2) DBLP articles are linked to Microsoft Academic Graph (MAG)'s topic tags, called field of study (FoS). The FoS tags were created by hierarchical topic modelling (Shen et al., 2018), with each article containing one or more FoS tags. We directly translated these well recognised FoS tags as topics, identifying 89,504 distinctive topics. On average, each topic was mentioned in around 54.68 papers as an indication of topic scale.

In addition to the entire DBLP dataset, we retrieved 11,399 articles on the information science (IS) disciplines from nine representative IS journals as defined by Hou et al. (2018) – i.e., *JASIST*, *Information Processing & Management*, *Journal of Informetrics*, *Information Research*, *Library & Information Science Research*, *Scientometrics*, *Research Evaluation*, *Journal of Documentation*, and *Journal of Information Science*. This sub-dataset contained 14,521 distinctive authors and 7,028 FoS tags and became our 'local' dataset.

*Experiment I: Local dataset (the sub-dataset for information science disciplines)*

Experiment I was designed to investigate the performance of the proposed method with the local dataset. This dataset contained a controllable number of articles with relatively high coupling, but a not-too-narrow topic list, as well as a general preference for research collaboration.

---
[1] https://dblp.org/

We divided the sub-dataset into two sets: Articles published in 2015 and before as the training set, and articles after 2015 as the testing set. We used authors, FoS tags, and their co-occurrences to build up the co-authorship layer and the co-topic layer, as well as edges connecting authors and topics. With these steps completed, we constructed two bi-layer networks, one for training purposes and the other for testing. The statistical information of the two networks is given in Table 1.

Table 1. Statistical information on the training and testing bi-layer networks (Experiment I)

|  | *Training network (2015 and before)* |  | *Testing network (After 2015)* |  |
|---|---|---|---|---|
|  | *# Nodes[1]* | *# Edges* | *# Nodes* | *# Edges* |
| Co-authorship layer | 11,836 | 15,510 | 3,415 | 5,525 |
| Co-topic layer | 6,497 | 122,531 | 2,348 | 24,226 |
| $E_{at}$ | 18,333 | 137,779 | 5,763 | 30,329 |
| # Papers | 9,908 (86.9%) |  | 1,491 (13.1%) |  |

Note: (1) # represents the number of related items.

Notably, there are 730 authors and 1817 FoS tags appearing in both networks, which thus becomes the key to the validation. In the training network, there are 1,326,410 possible edges between the 730 target authors and the 1817 topics. With 14,184 existing edges, our key task was to predict and rank the 1,312,226 non-existent edges and match the top-ranking ones with the 30,329 'true' edges in the testing network.

Targeting four groups of the top-ranking edges (i.e., the overall performance for the 1,312,226 edges, and the performance for the top 500, 1000, and 1500 edges with highest prediction scores, respectively), the ROC curves and AUC values of the proposed method (noted as "diffusion") and the seven baselines are given in Figure 2.

The proposed method demonstrates a recognisable advantage compared to its counterparts across all four groups, and it is particularly superior in recommending the top-ranked topics for target users. This well matches the preference of recommendation in practice, i.e., the higher the ranking an item has, the better it meets actual user needs.

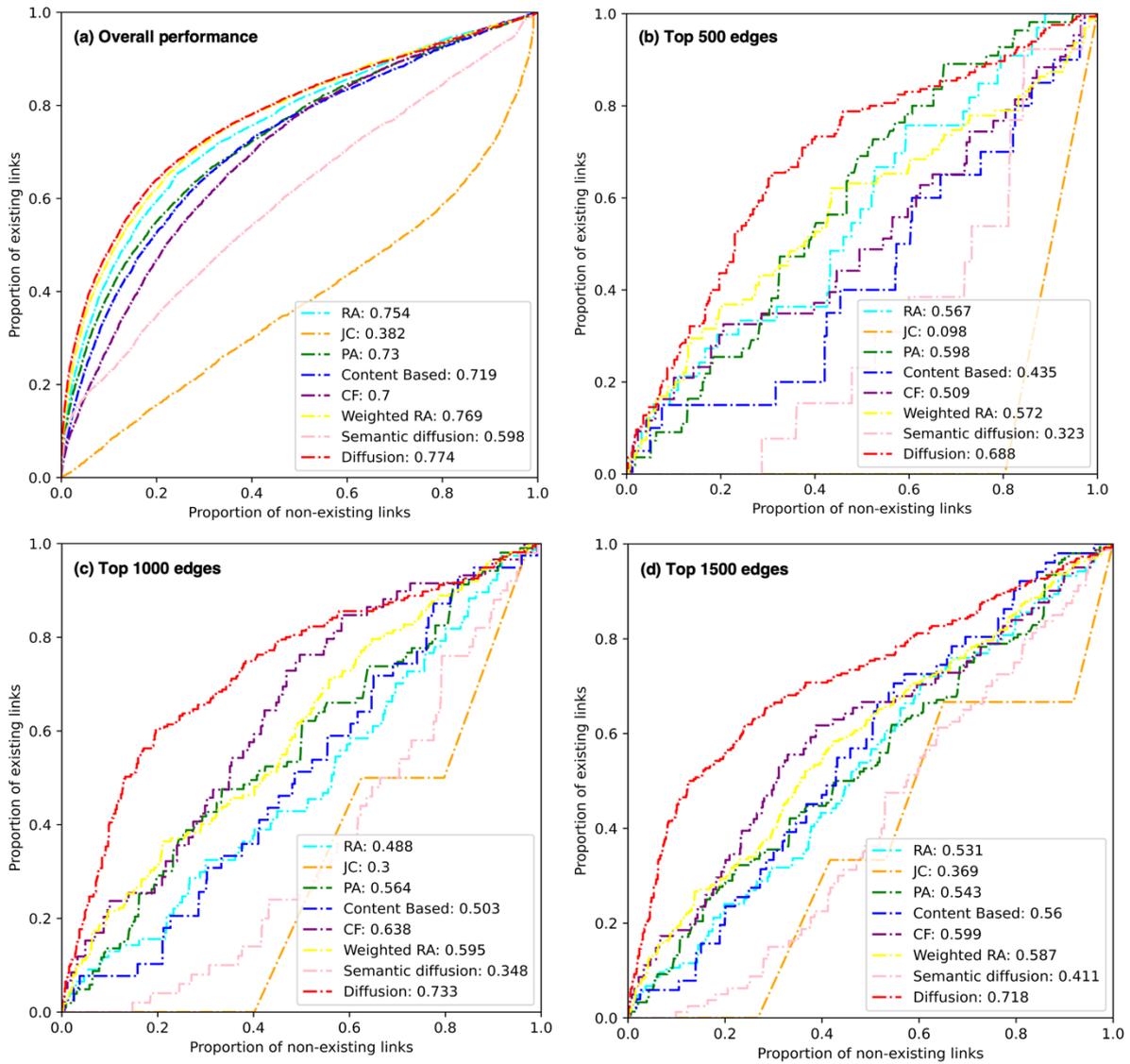

Figure 2. AUC curves of the proposed method and the six baselines (Experiment I)

We made certain further interpretations in terms of this comparison. (1) The topological structure of our bi-layer bibliometric network is extremely imbalanced. Imagining a productive researcher with active collaborations that spans a broad range of research topics vs a rigorous researcher with limited collaborations who only concentrates on a few topics. The proposed method specifies two types of common neighbours (i.e., co-authors and co-occurrent topics). According to the results, particularly those for recommending top-ranked topics in Figures 3(b-d), the proposed method significantly increases the prediction performance, compared to those CN-based baselines. (2) In terms of the two recommender system baselines, the content-based approach simply uses the co-occurrence of topics. The CF-based approach is slightly similar to the NBI approach with a bipartite network in that it uses one set of nodes (e.g., authors) to represent the other set of nodes (e.g., topics). As claimed, our method well differentiates the two sets of heterogeneous nodes and three types of interactions between them, and such efforts have gained a return in improving the performance (3) Interestingly, the unpreferred but expected performance of the semantic diffusion baseline demonstrates the drawback of using embedding techniques with a local bibliometric dataset. Compared to a large sparse co-topic layer, a well-connected semantic layer might introduce a great deal of noise.

*Experiment II: Global dataset (the DBLP dataset)*

Experiment II was designed to examine the proposed method with a large-scale dataset that covers distinct research topics. Although the DBLP data focuses on computer science and may not be a typical global dataset, we argue that the rapid development of information technology over past decades has led to active disciplinary interactions that cross relatively broad and diverse research areas. In fact, the dataset spans seven of the Web of Science research areas[2]: artificial intelligence, cybernetics, information systems, software engineering, theory and methods, hardware and architecture, and interdisciplinary applications. Thus, the DBLP data can be thought of as a 'not-bad' example of a global dataset for the purposes of our experiments.

We followed the same phases of data pre-processing and network construction as for Experiment I. Table 2 provides the statistical information of the two bi-layer networks.

Table 2. Statistical information of the training and testing bi-layer networks (Experiment II)

|  | *Training network (2015 and before)* | | *Testing network (After 2015)* | |
| --- | --- | --- | --- | --- |
|  | # Node | # Edge | # Node | # Edge |
| Co-authorship layer | 3,173,445 | 9,089,406 | 1,755,417 | 5,746,287 |
| Co-topic layer | 83,563 | 13,493,950 | 65,904 | 6,757,798 |
| $E_{at}$ | 3,257,008 | 56,547,787 | 1,758,191 | 26,175,654 |
| # Papers | 3,610,096 (73.8%) | | 1,283,985 (26.2%) | |

Ranking these million edges was time-consuming and required a great deal of computational power. Thus, we followed an efficient measurement for large-scale network analytics designed by Zhou et al. (2009). More specifically, we created a probe set containing 20,000 randomly selected author-topic edges $E_{at}$. Not all of these edges exist in the training network, and only half of them appear in the testing network. These are the true labels. The other half do not, and these are the false labels. We then validated the results according to this

---

[2] https://incites.help.clarivate.com/Content/Research-Areas/wos-research-areas.htm

'randomly selected' probe set. The average AUC curves of the proposed method and the six baselines are given in Figure 3.

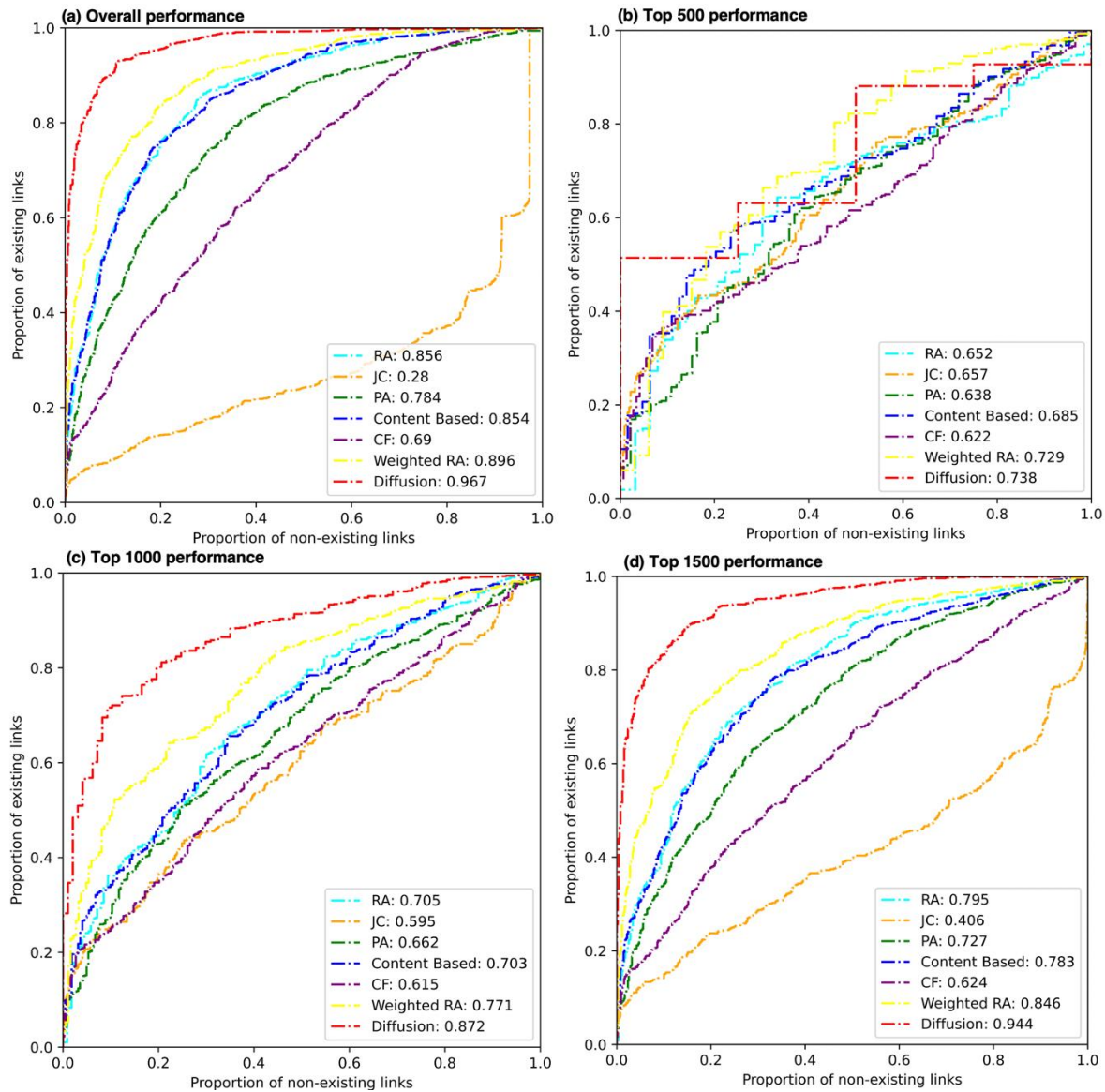

Figure 3. AUC curves of the proposed method and the six baselines (Experiment II)

Figure 3 demonstrates a prior performance of the proposed method in this global dataset. Specifically, the proposed method shows a distinct advantage in overall performance but its superior with the weighted RA approach was narrowed in recommending the 500 top-ranked topics. The comparison between the proposed method and the baselines generally follows a similar trend to that of Experiment I. However, we noted a relatively high AUC value (c.f., 0.97) for Experiment II. This is because Experiment I needed to identify 30,329 edges from 1.3 million, while Experiment II involved the much easier task of identifying 10,000 true labels from 20,000 edges. Additionally, we skipped off the semantic diffusion approach in

Experiment II due to its computational complexity, which requires to calculate the similarities between all the 34.9 million pairs with the 83,563 topics.

*Case study: Recommending knowledge trajectories for information scientists*

To further verify the merits of the method, we conducted a case study with the local dataset of information science articles. Our goal was to chart the knowledge trajectories of the researchers in the dataset. The results of the study not only showcase the qualitative substance of the proposed method, they also demonstrate a practical way to support decision-making for individual researchers, policymakers, and entrepreneurs.

Different to Experiment I, in the case study, we used the entire local dataset to construct the bi-layer network, which included 14,521 authors and 7,028 FoS tags. Our key task was to recommend knowledge trajectories for these 14,521 authors beyond their current foci. The statistical information of the bi-layer network is given in Table 3.

Table 3. Statistical information of the bi-layer networks (case study)

|  | *# Node* | *# Edge* |
| --- | --- | --- |
| Co-authorship layer | 14,521 | 20,704 |
| Co-topic layer | 7,028 | 137,088 |
| $E_{at}$ | 21,549 | 165,222 |
| Total | 21,549 | 536,299 |

Diffusion-based prediction was applied to recommend the FoS tags as research topics and 100 topics were recommended for each author. The full list of this recommendation can be reviewed in the Table S1[3]. However, to better demonstrate the recommendations for the purposes of this article, we applied the Leiden approach of community detection (Traag et al., 2019) to identify author and topic communities, and thus transformed our task into recommending topics to target author communities. Following the default parameters of the Leiden approach, we identified 4,352 author communities (see Table S2) and 8 topic communities (see Table S3). The descriptive statistics of their community sizes is given in Table 4.

Table 4. Descriptive statistics of the community sizes of author and topic communities

---
[3] We created a GitHub repository for Supplementary Documents, Tables S1-S3 can be retrieved from the link: https://github.com/IntelligentBibliometrics/Diffusion

|         | *Max*  | *Min* | *Average* | *Standard Deviation* |
|---------|--------|-------|-----------|----------------------|
| Author  | 337    | 1     | 3.34      | 12.50                |
| Topic   | 1604   | 14    | 878.5     | 577.14               |

Despite a relatively large amount of small author communities that only contained one author, we screened the top authors in ten of the largest communities and observed coherent and promising clustering strategies (see Table 5). The ten communities mainly group researchers based on geographical feature (e.g., US, UK, Europe, and Tampere University) and research interests (e.g., information retrieval and information systems), coinciding with the general motivation and diving forces of research collaborations. We labelled these ten communities based on the main composition of their representative authors. Interestingly, we observed three bibliometric communities among the top ranks:

- Community 1: US-based and mainly relates to the Indiana University Bloomington.
- Community 2: Europe-based and includes well-recognised bibliometricians covering broad bibliometric topics.
- Community 5: Includes bibliometricians in research evaluation, such as h-index and impact factors, and particularly some pioneer researchers.

Table 5. Representative authors of top ten author communities

|   | *Label*                        | *Size* | *Representative authors* |
|---|--------------------------------|--------|--------------------------|
| 1 | Indiana U-lead bibliometrics   | 337    | Ying Ding, Blaise Cronin, Erjia Yan, Vincent Larivière, Robert M. Losee, Cassidy R. Sugimoto, Min Song, Schubert Foo |
| 2 | EU-lead bibliometrics          | 286    | Lutz Bornmann, Loet Leydesdorff, Ludo Waltman, Hans-Dieter Daniel, Félix de Moya-Anegón, Nees Jan van Eck, Henk F. Moed, Rüdiger Mutz |
| 3 | UK-lead IS                     | 226    | Charles Oppenheim, Peter Willett, David Bawden, Nigel Ford, Reijo Savolainen, Lyn Robinson, F. E. Wood, Paul Clough |
| 4 | US-lead IS                     | 225    | Donald H. Kraft, Tefko Saracevic, Katherine W. McCain, James Hartley, Gary Marchionini, Chirag Shah, Harold Borko, Edward A. Fox |
| 5 | H-index                        | 204    | Ronald Rousseau, Leo Egghe, Abraham Bookstein, Quentin L. Burrell, Fred Y. Ye, Tove Faber Frandsen, Raf Guns, Per Ahlgren |
| 6 | ISys | 187 | Hsinchun Chen, Christopher C. Yang, Michael Chau, Bruce R. Schatz, Chih-Ping Wei, Wingyan Chung, Guang Yu, Lina Zhou |
| 7 | IR | 185 | Mounia Lalmas, Fazli Can, Joemon M. Jose, Aixin Sun, Berkant Barla Cambazoglu, Ryen W. White, C.J. Van Rijsbergen, William B. Rouse |
| 8 | US-lead ISys & HCI | 176 | Paul Kantor, Nicholas J. Belkin, W. Bruce Croft, Bracha Shapira, Diane Kelly, Xiangmin Zhang, Jacek Gwizdka, R. W. P. Luk |
| 9 | Tampere U-lead IR | 156 | Kalervo Järvelin, Judit Bar-Ilan, Pertti Vakkari, Mark Sanderson, Timo Niemi, Bluma C. Peritz, Soo Young Rieh, Preben Hansen |
| 10 | UK/EU-lead IR | 152 | Ian Ruthven, Fabio Crestani, Maarten de Rijke, Shengli Wu, Massimo Melucci, Dawei Song, P. D. Bruza, Nicola Ferro |

Note: IS = information science; IR = information retrieval; ISys = information systems; HCI = human-computer interaction.

We also manually checked the eight topic communities and combined the two smallest communities with less than 100 topics into their most similar communities. The final six communities and their representative topics are given in Table 6. The #1 "ranking" and #5 "information retrieval" communities include the major artificial intelligence and data science techniques involved in information studies. #3 "data mining & bibliometrics" highlights the role of bibliometrics and its close connections with data analytics. The other three communities hold their relatively unique emphases, e.g., information systems and knowledge management, information seeking and social media, and digital libraries.

Table 6. The six topic communities and representative topics

| | *Label* | *Size* | *Representative topics* |
|---|---|---|---|
| 1 | Ranking | 1618 | natural language processing, search engine indexing, document retrieval, machine learning, cluster analysis |
| 2 | Information systems | 1471 | knowledge management, social science, operations research, public relations, information access |
| 3 | Data mining & bibliometrics | 1181 | data science, citation analysis, artificial intelligence, scientometrics, statistics |
| 4 | World wide web & Information seeking | 1202 | information needs, social network, multimedia, content analysis, cognition |
| 5 | Information retrieval | 1115 | computer graphics, faceted classification, music information retrieval, computer network, information security |
| 6 | Digital library & ontology | 441 | metadata, cataloguing, document structure description, semantic web, database design |

Note that we labelled topic communities based on their highest frequent terms, but we added some supportive terms to some communities for a comprehensive coverage.

To map these author and topic communities, we used Gephi's OpenOrd layout (see Figures 4 and 5). Note that, in Figure 4, we only coloured the top 1000 largest author communities; the remaining communities appear in grey. Generally, numerous author communities are scattered across the map, which indicates relative independence between the groups. However, in Figure 5, the topic communities are closely aggregated together, reflecting close interdisciplinary interactions within the IS discipline.

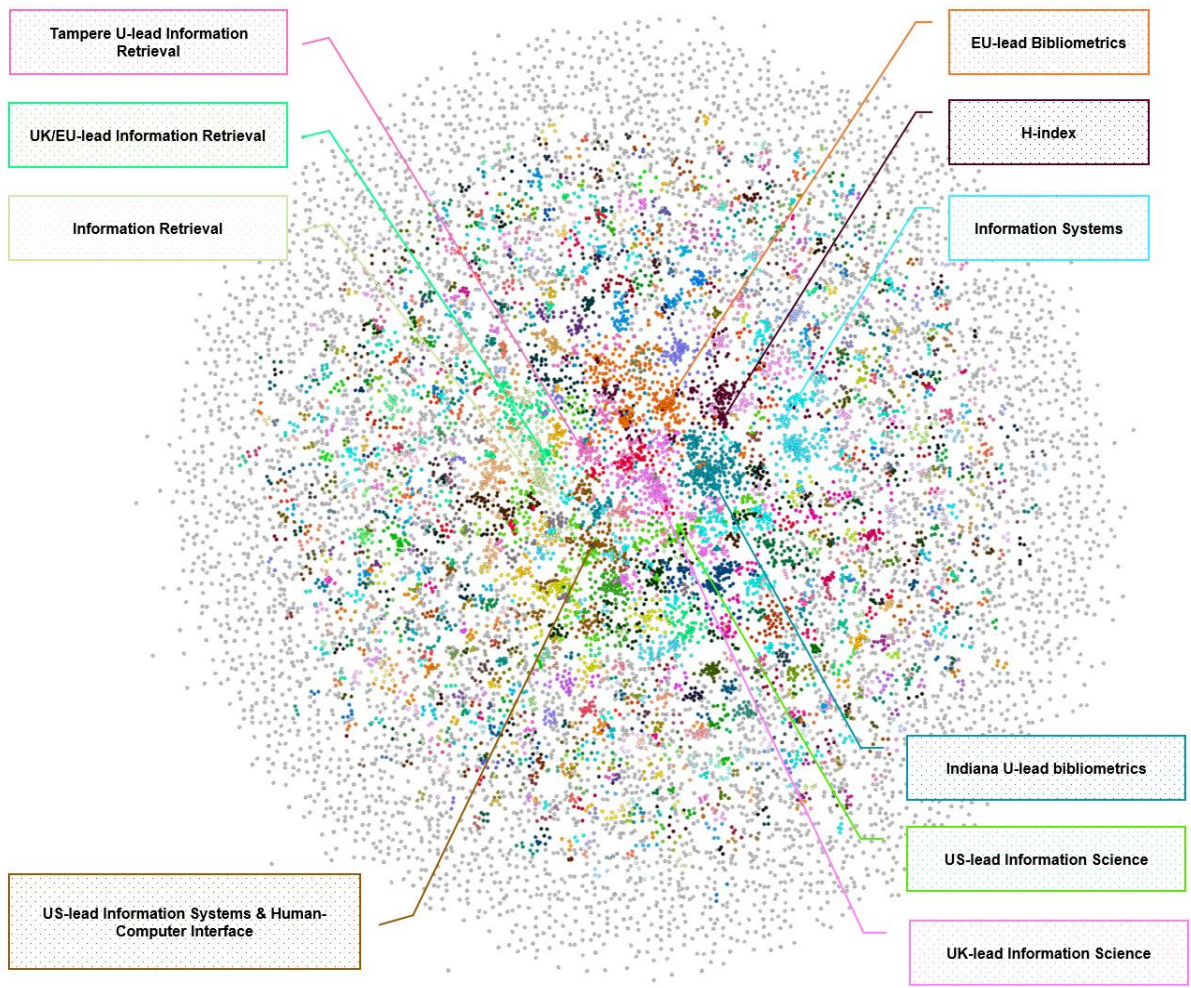

Figure 4. Author communities of the information science discipline

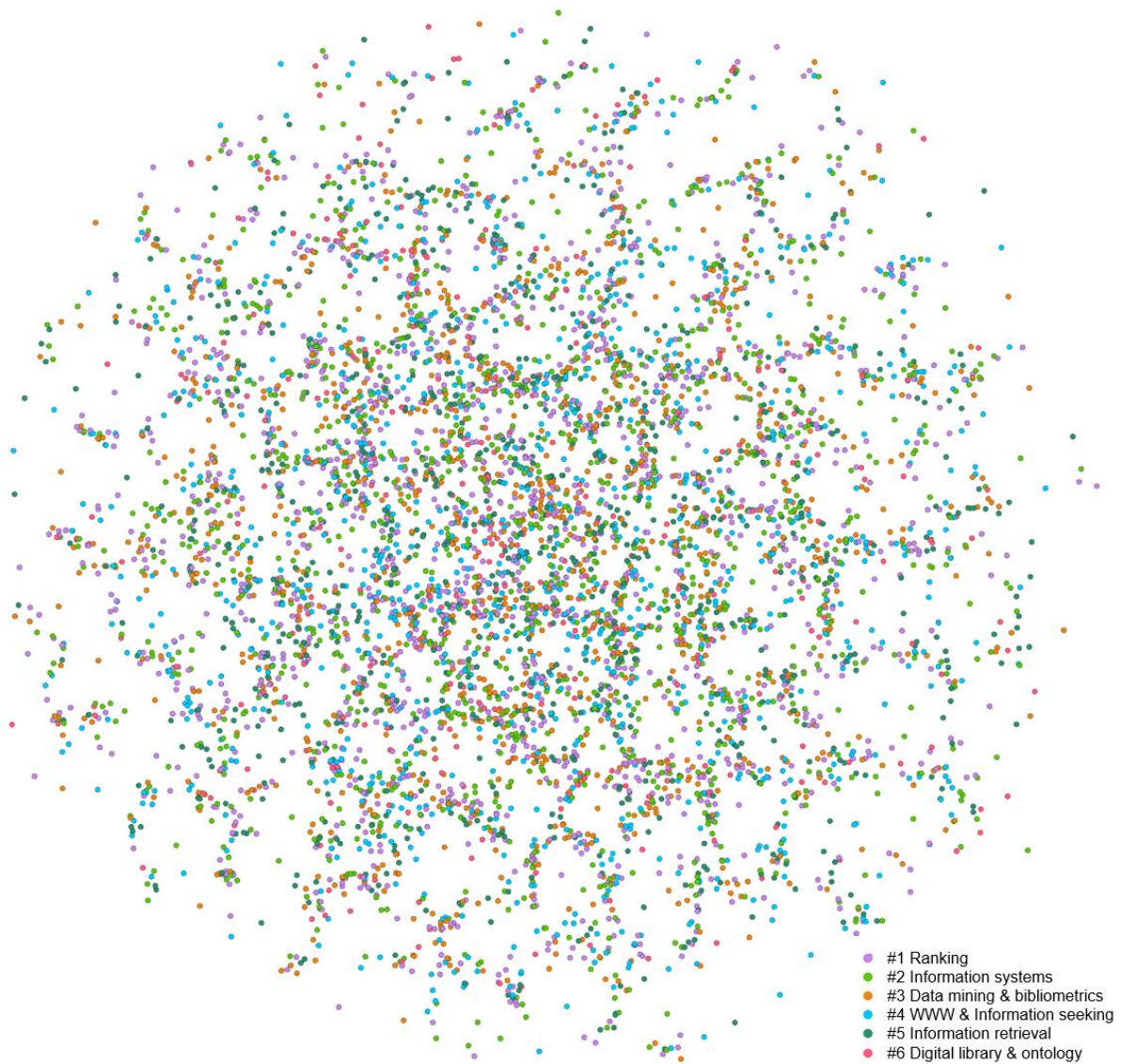

Figure 5. Topic communities of the information science discipline

Taking the top ten author communities, we generated recommendations of research topics to the authors involved and summed up individual topics to create mapping between author and topic communities. Figure 6 reveals the distribution of recommended topics to these top ten author communities. One trend we observed was that the larger the topic community, the higher the chance it had to be recommended. However, since we observed relatively few "information retrieval" topics were recommended, we might interpret as its wide involvement in information studies and our method recommends topics beyond the current foci of the target authors. Similarly, "data mining & bibliometrics" is also a community in such a situation, compared to a much larger number of recommended topics from a similar-size community "world wide web & information seeking".

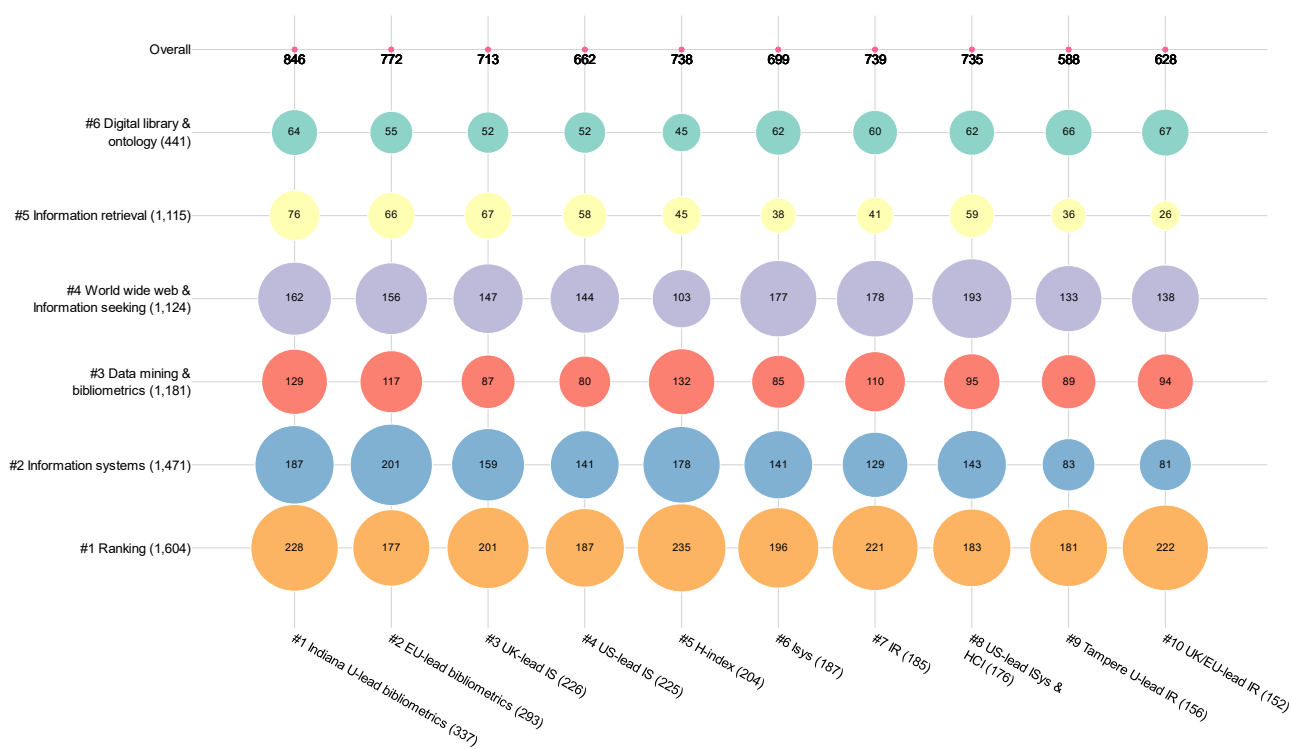

Figure 6. Distribution of topic recommendation for targeting author communities

Note that the numbers after the community labels indicate the size of the related community. The overall line at the top of the figure indicates the total number of topics recommended to the related author communities.

Table 7 lists the top recommended topics for the top ten author communities. Despite being the largest community, author community #1 has received some disruptive recommendations. For example, "knowledge base" was recommended to 57 of the 337 authors in the community, while "information needs" was recommended to 254 of the 286 authors in author community #2. Given the cohort of IS topics, Community 1 has touched on much broader topic landscapes than that of Community 2. This could be interpreted by the expertise of representative authors in these communities. While the authors in Community 2 mainly focus on bibliometrics, those in Community 1 might reach out into technical base and practical applications of bibliometrics. For instance, Prof Ying Ding has been closely involved in computer science research (Ding et al., 2021), Prof Cassidy Sugimoto is closely with public policy (Kozlowski et al., 2022), and Prof Min Song is an expert in bioinformatics (Hong et al., 2021). Thus, we might anticipate a follow-up study using the distribution of recommended topics to measure the topic disruption of research communities. Besides their current foci, applying indicators to these recommended topics, such as entropy or the Herfindahl–

Hirschman index, could be a secondary measure to understand the diversity of their research interests.

In terms of an empirical evaluation, for Community 1, it is easy to recognise that topics "knowledge base", "information behaviours", and "personalisation" could be some follow-up steps for information studies, e.g., behaviour analysis and recommender systems. In terms of Community 2, those IR-related topics could be well endorsed by some urgent bibliometric trends, e.g., scientific IR (Waltman, 2021). By contrast, Community 7 is already heavily involved in IR, so the topics "bibliometrics", "social science", and "scientometrics" were recommended. This aligns with increasing interest from the IR community in bridging bibliometrics with IR, e.g., bibliometrics-enhanced IR (Mayr et al., 2014). "Machine learning" was recommended to Community 3, most of whom are UK-based information scientists. Interestingly, this group mainly focuses on information studies related to the social sciences and humanities, such as digital literacy (Bawden, 2001) and information law (Oppenheim et al., 2020). Thus, the future use of intelligent information technologies might create complementary value.

Table 7. Top recommended topics for the top ten author communities

| Author Community | Top 10 Topics |
| --- | --- |
| 1 - Indiana U-lead bibliometrics | knowledge base (57), query language (55), information behaviour (42), law (36), pedagogy (26), personalisation (23), personal information management (22), China (22), web modelling (21), microblogging (19) |
| 2 - EU-lead bibliometrics | information needs (254), user interface (120), cognitive models of information retrieval (57), human–computer information retrieval (54), ranking (information retrieval) (50), management information systems (49), recall (49), automatic summarisation (46), usability (45), auteur theory (41) |
| 3 - UK-lead IS | machine learning (85), vocabulary (73), recall (59), statistics (58), syntax (43), sentence (40), citation impact (34), developing country (33), phrase (33), parsing (28) |
| 4 - US-lead IS | impact factor (93), categorisation (68), hyperlink (51), syntax (50), internet privacy (48), similitude (43), hypertext (40), classifier (linguistics) (37), information and communications technology (37), government (34) |
| 5 - H-index | query expansion (132), digital library (114), user interface (99), information access (86), information dissemination (67), electronic publishing (62), information management (58), social media (49), ranking (information retrieval) (44), systems design (42) |
| 6 - ISys | ranking (183), library science (162), database (148), cognition (107), social science (104), weighting (93), sociology (88), scientific literature (72), public relations (63), higher education (58) |
| 7 - IR | bibliometrics (164), social science (103), scientometrics (85), public relations (81), scientific method (76), scientific literature (69), impact factor (62), sentiment analysis (61), information management (60), statistics (50) |
| 8 - US-lead ISys & HCI | social network (122), social science (117), public relations (93), scientific method (77), scientometrics (67), automatic summarisation (63), scientific literature (62), sentiment analysis (55), social media (51), end user (46) |
| 9 - Tampere U-lead IR | cluster analysis (116), probabilistic logic (57), knowledge base (53), algorithm (49), vector space model (44), parsing (41), web mining (33), concept search (30), ontology (information science) (30), epistemology (28) |
| 10 - UK/EU-lead IR | citation analysis (146), bibliometrics (135), information technology (116), natural language (115), operations research (105), information processing (99), automatic indexing (96), social science (79), software (78), artificial intelligence (78) |

Note that the number within the brackets represents the number of authors in this community who were recommended this topic.

**Discussion and conclusions**

In this study we developed a method of diffusion-based network analytics for recommending knowledge trajectories. We constructed a heterogeneous bibliometric network consisting of a co-authorship layer and a co-topic layer, analysed the process of knowledge diffusion between authors and research topics, and recommended personalised topics for target authors that lie beyond their current research foci and could be their future knowledge trajectories.

*Technical and practical implications*

This method of diffusion-based network analytics is non-parametric, which does not contain any super parameters requiring human intervention or extra experiments. However, several sensitive factors that might influence the practical use of the method exist:

*Local dataset vs. global dataset*: Technically, the proposed method requires exhaustive calculations for all possible edges in a network. Thus, when the scale of the heterogeneous network increases, the growth of the complexity could be exponential, and the method might become extremely time-consuming.

*Topic models* for constructing the co-topic layer: This study directly facilitated MAG's FoS tags. However, in practical use, selecting an appropriate approach for topic extraction could be an issue. Traditional bibliometric studies often rely on clustering algorithms like K-means (Zhang, Lu, et al., 2018) and principal component analysis (PCA) (Wang et al., 2018) to set the scale of topics to a controllable level, whereas our method prefers a relatively large number of topics to maintain the topology of the co-topic layer. One alternative solution is to directly use author keywords or terms retrieved from the combined titles and abstracts for topic extraction.

The selection of *top-ranked topics*: Experiment I provides a clear clue to demonstrate the prior performance of the proposed method in top-ranked topics, since in practical use, we may only need several topics for recommendation, rather than hundreds or thousands topics. Our case study has created a feasible way to use author communities and topic communities and recommend a readable number of topics in empirical cases.

*Limitations and future directions*

As with all studies, ours has limitations that provide opportunities for future research. (1) To process large-scale global datasets, we may need to investigate distributed systems or parallel computing techniques. These may provide a more efficient solution than the one presented. (2) Despite acceptable reasons for using the DBLP data as a global dataset, applying the proposed method to some well-recognised global datasets (e.g., MAG, Web of Science, and Scopus) may not only enhance technical benefits but also create practical significance, such as understanding multidisciplinary interactions. (3) One interesting follow-up direction to this study includes introducing dynamic network analytics to capture the cumulative changes over time when knowledge trajectories emphasise the dynamic process of scientific and technological evolution.

**Acknowledgements**

This work is supported by the Australian Research Council under Discovery Early Career Researcher Award DE190100994.